\documentclass[twocolumn, a4paper,fleqn,usenatbib]{aastex6}    
\usepackage[T1]{fontenc}
\usepackage{ae,aecompl}

\usepackage{graphicx}	
\usepackage{amsmath}	
\usepackage{amssymb}	
\usepackage{bm}
\usepackage{subfig}
\usepackage{times}

\newcommand{\ie}{i.e.,~}
\newcommand{\eg}{e.g.,~}

\newcommand{\bhac}{\texttt{BHAC}~}

\begin{document}

\title{Magnetically inspired explosive outflows from neutron-star
  mergers}

\author{Antonios Nathanail\altaffilmark{1}, Oliver
  Porth\altaffilmark{2,1}, Luciano Rezzolla\altaffilmark{1}}
\altaffiltext{1}{Institut f\"ur Theoretische Physik, Max-von-Laue-Strasse
  1, D-60438 Frankfurt, Germany}
\altaffiltext{2}{Astronomical Institute Anton Pannekoek, University of
  Amsterdam, Science Park 904, 1098 XH, Amsterdam, The Netherlands}
\begin{abstract}
Binary neutron-star mergers have long been associated with short-duration
gamma-ray bursts (GRBs). This connection was confirmed with the first
coincident detection of gravitational waves together with electromagnetic
radiation from GW170817. The basic paradigm for short-duration GRBs
includes an ultra-relativistic jet, but the low-luminosity prompt
emission together with follow-up radio and X-ray observations have hinted
that this picture may be different in the case of GW170817. In
particular, it has been proposed that large amounts of the magnetic
energy that is amplified after the merger, can be released when the
remnant collapses to a black hole, giving rise to a quasi-spherical
explosion impacting on the merger ejecta. Through numerical simulations
we investigate this scenario for a range of viewing angles, injected
energies and matter densities at the time of the collapse.  
Depending on the magnitude of the energy injection and the remnant density, 
we find two types of outflows: one with a narrow relativistic core and one with a wide-angle, 
but mildly relativistic outflow. Furthermore, very wide outflows are
possible, but require energy releases in excess of
$10^{52}\,\mathrm{erg}$.
\end{abstract}

\section{Introduction} 
\label{sec:intro}
The coincident and follow-up observations across the electromagnetic
spectrum of the gravitational waves from the event
GW170817~\citep{Abbott2017_etal,Abbott2017b_etal} have 
introduced a new era for the study of
dense matter.   
 The detection of GRB 170817A, \ie,
the electromagnetic counterpart to GW170817, has established the
connection between binary neutron-star (BNS) mergers and short gamma-ray
bursts (GRBs).    
Models have predicted such a connection
through the production of a jet following the
merger~\citep{Eichler89,Narayan92,Rezzolla:2011}.

GRB170817A was a subluminous event suggesting that we were possibly not
looking straight into a jet~\citep{Goldstein2017,Savchenko2017}. The
prompt emission alone could not give a clear picture of whether this is
intrinsic or due to other factors, one possibility being the large
observing angle with respect to the axis of the outflow that produced the
emission. Furthermore, the first radio and X-ray afterglow detection tens
of days after merger, could not clarify the
picture ~\citep{Hallinan2017,Margutti2017,Troja2017}. Subsequent radio
and X-ray observations have provided important clues, but it is still
difficult to clearly distinguish between different models for the
emission~\citep{Alexander2017,Alexander2018,DAvanzo2018,
Dobie2018,Margutti2018,Mooley2018,Nynka2018,Troja2018}. A
hundred days after the main event, the radio flux was still rising and
showed that the outflow has somehow a radial and/or angular structure.
However, the most recent observations favor a rather collimated outflow
that possess a relativistic core with an opening angle $< 5^{\circ}$
\citep{Mooley2018b,Mooley2018c}.

Some of the proposed scenarios invoke the successful launch of a
relativistic jet; however it is not seen directly and only
the cocoon is visible. Another possibility is that the jet may not break
out from the BNS ejecta, but merely deposit its energy there. Detailed
discussions on models discussing either off-axis radiation from a jet or
the cocoon emission from a choked jet have been presented
by~\citet{Murguia-Berthier2016,Murguia-Berthier2017,Lazzati2017a,
Gottlieb2018,Bromberg2018,Kathirgamaraju2018,
Hotokezaka2018,Xie2018,Salafia2018a}.

General-relativistic simulations have recently drawn a robust picture of
BNS mergers, showing that mass ejection can be significant during merger,
from which a visible kilonova is produced due to \textit{r}-processes. The
standard short-GRB picture envisions a jet to be produced after the
collapse to a black hole of the merger remnant.  If a jet was produced in
GRB 170817A, its interaction with the ejecta is probably at the heart of
the phenomenology observed.

Despite the differences in the numerous proposed models, a shared
featured is the assumption that shortly after merger the remnant
collapses to a black hole~
\citep{Granot2017,Margalit2017,Shibata2017c,Metzger2018,Nathanail2018,
Rezzolla2017}.
In this paper we describe a different possibility that naturally gives
rise to two families of outflows following the merger: one with a narrow
relativistic core and one with a wide-angle, mildly relativistic outflow.
More specifically, depending on the collapse time and properties of the
surrounding matter, we show that a jet need not be
launched ~\citep{Lehner2011,Nathanail2018}. However, the collapse of the
remnant could initiate an ``explosion,'' namely, a sudden release of
large amounts of the magnetic energy that was amplified via instabilities
during the early stages of the merger, and that can reach values even in
excess of $10^{51}\,\textrm{erg}$~\citep{Kiuchi2017}. When this energy is
released, it can push and accelerate the material in the vicinity of the
compact remnant. This accelerated material will not be able to propagate
through the high-density torus, which is mostly concentrated on the
binary orbital plane. Simulations also show that the density of the
material material around the compact remnant has a strong angular
dependence, being considerably smaller toward the orbital
axis ~\citep{Bovard2017}. As a result, the produced outflow has a natural
``escape route'' along the polar direction, where it can propagate
essentially freely and where, when accelerated, can quickly catch up to
the slow-moving ejecta~\citep{Rezzolla2014b}.

Several hydrodynamic simulations in multidimensions have modeled an
injected outflow that passes through the BNS
ejecta~\citep{Lazzati2017a,Murguia-Berthier2017,Gottlieb2018,Kathirgamaraju2018,Xie2018}.
Here, on the other hand, the outflow is the product of a \emph{single}
spherical energy release (explosion) triggered by the collapse of the
compact remnant. Hence, all of the energy is injected at one instant in
time and not continuously; this represents an important difference with
respect to previous work and leads, indeed, to novel features.

In particular, we investigate such a scenario through general-relativistic
two-dimensional simulations where the BNS ejecta are described in terms
of a torus whose size and density profile depend on the time when the
compact remnant of the merger collapses to a black hole, with the the
maximum density of the torus $\rho_{\mathrm{tor}}$ decreasing as the time
of collapse $t_{\mathrm{coll}}$ is increased. We then vary the amount of
energy injected at the time of collapse in the range of
$10^{48}\,-\,10^{52}\,\mathrm{erg}$ and find that the energy injection
leads quite robustly to either a wide-angle, mildly relativistic outflow
or an outflow with a relativistic and narrow core. Note that while an
``unsuccessful jet'' normally refers to a jet that is produced but cannot
emerge through the BNS ejecta, a jet is never produced in our study.

Particular attention should be paid to our models \texttt{\small{E.50.9}}
and \texttt{\small{E.51.10}} which have a faster-moving core in a cone
$\lesssim10^{\circ}$ and reach $\Gamma\gtrsim10$ (models
\texttt{\small{E.51.9}} and \texttt{\small{E.52.10}} have also a
faster-moving core, but a slightly different energy distribution). These
properties make them compatible with the observational constraints
reported in \cite{Mooley2018b}, who argue that the emission around the
peak of the lightcurve come from a component with angular extent
$\lesssim14^{\circ}$ and $\Gamma\simeq4$. Furthermore, because of the
faster-moving core, these models should naturally lead to a steep
post-peak decay, again in agreement with the interpretations of
\cite{Mooley2018c}.

\begin{table}
  \centering
    \caption{Properties of the Various Scenarios Considered: Energy Released
    at Collapse $E_{\mathrm{exp}}$, Maximum Density of the Torus
    $\rho_{\mathrm{tor}}$, Time of Collapse $t_{\mathrm{coll}}$, Average
    Lorentz Factor $\langle\Gamma\rangle$, Averaged Lorentz Factor
    $\langle\Gamma\rangle_{30}$ within $30^{\circ}$ from the Polar Axis,
    and the Outflow Mass Moving with $\Gamma>1.2$.}
  \begin{tabular}{l||c|c|c|c|c|r}
    \hline
    \hline
    \!\!\!  model & $E_{\mathrm{exp}}$&$\rho_{\mathrm{tor}}$&$t_{\mathrm{coll}}$& 
    \!\!\! $\langle\Gamma\rangle$&$\langle\Gamma\rangle_{30}$&$M_{(\Gamma>1.2)}$  \\
    \!\!\! &$(\mathrm{erg})$&$(\mathrm{g/cm^3})$  &
    \!\!\! $(\mathrm{sec})$& & &$(10^{-7}M_{\odot})$\\
    \hline
    \! \texttt{\small{E.50.11}} & $10^{50}$ &$10^{11}$& $0.5$ & $1.10$ & $1.21$ & $6$    \\
    \! \texttt{\small{E.51.11}} & $10^{51}$ &$10^{11}$& $0.5$ & $1.50$ & $2.48$ & $90$   \\
    \! \texttt{\small{E.52.11}} & $10^{52}$ &$10^{11}$& $0.5$ & $2.53$ & $5.57$ & $630$  \\
    \! \texttt{\small{E.49.10}} & $10^{49}$ &$10^{10}$& $1.0$ & $1.10$ & $1.21$ & $0.6$  \\
    \! \texttt{\small{E.50.10}} & $10^{50}$ &$10^{10}$& $1.0$ & $1.50$ & $2.48$ & $9$    \\
    \! \texttt{\small{E.51.10}} & $10^{51}$ &$10^{10}$& $1.0$ & $2.53$ & $5.57$ & $63$   \\
    \! \texttt{\small{E.52.10}} & $10^{52}$ &$10^{10}$& $1.0$ & $2.39$ & $4.47$ & $1400$ \\
    \! \texttt{\small{E.48.09}} & $10^{48}$ &$10^{9}$ & $2.0$ & $1.10$ & $1.21$ & $0.06$ \\
    \! \texttt{\small{E.49.09}} & $10^{49}$ &$10^{9}$ & $2.0$ & $1.50$ & $2.48$ & $0.9$  \\
    \! \texttt{\small{E.50.09}} & $10^{50}$ &$10^{9}$ & $2.0$ & $2.53$ & $5.57$ & $6.3$  \\
    \! \texttt{\small{E.51.09}} & $10^{51}$ &$10^{9}$ & $2.0$ & $2.39$ & $4.47$ & $140$  \\
    \! \texttt{\small{E.52.09}} & $10^{52}$ &$10^{9}$ & $2.0$ & $3.26$ & $4.55$ & $3000$ \\
   \hline
    \hline
  \end{tabular}
  \label{tab:initial}
\end{table}

\section{Numerical setup} 
\label{sec:num}
We employ \bhac to solve the relativistic-hydrodynamic equations in a
Kerr background spacetime~\citep{Porth2017}. The initial setup is chosen
to resemble the ejected matter and the expanded torus around the compact
remnant that formed during the BNS merger. Our simulations are performed
in two spatial dimensions exploiting the approximate azimuthal symmetry
of the system and we take advantage of three levels of mesh refinement to
resolve the outflow at an effective resolution of $2048\times512$ cells
(the radial grid has a logarithmic spacing with a minimum size of
$40\,\mathrm{m}$). We initialize the fluid with an equilibrium torus with
constant specific angular momentum~\citep{Fishbone76} around a Kerr black
hole with $M\sim2.7\,M_{\odot}$ and a dimensionless spin of
$a:=J/M^2=0.93$; although the latter is somewhat larger than what is
expected, the precise value used for the angular momentum $J$ has little
influence on our results because the energy injection is extremely rapid
and most of the fluid dynamics takes place far from the black hole. The
torus has a size of $1,200\,\mathrm{km}$ and is contained in domain of
radius $10,000\,\mathrm{km}$.

The parameters of the torus are chosen after considering that the remnant
of GW170817 must have survived $\lesssim\,1\,{\rm s}$ 
~\citep{Granot2017,Margalit2017,Shibata2017c,
Nathanail2018,Rezzolla2017}. Simulations
indicate that after its formation, the torus around the compact remnant
loses mass due to accretion, dynamical ejection via shock heating, and
the emission of winds driven either by magnetic fields or neutrinos. All
of these processes lead to redistribution of angular momentum which is
generically transported outward ~\citep{Rezzolla:2010}. Hence, the mass
and the maximum density of the torus must decreases over time leading us
to initial configurations with maximum densities of
$\rho_{\mathrm{tor}}\simeq10^{11},\,10^{10},\,10^9\,\mathrm{g/cm^3}$,
thus corresponding to collapse times of
$t_{\mathrm{coll}}\simeq0.5,\,1.0,\,2.0\,\mathrm{s}$, respectively; these
values match well the results of numerical
simulations~\citep{Bovard2017,Fujibayashi2017b}.

In a BNS merger, the region near the polar axis is filled with ejected
matter, although at lower densities than near the equatorial plane. To
reproduce these conditions together with the use of non-self-gravitating
equilibrium tori, we also fill this region with matter, as it has a density
that is two and a half orders of magnitude smaller than the maximum
density in the equilibrium torus and has a radial fall-off that scale
like $r^{-1.5}$, with $r$ being the radial distance from the black hole.
Overall, in our simulations the integrated rest-mass outside the black
hole is $M_{\mathrm{tor}}=6\times10^{-2}\,M_{\odot}$ and $
M_{\mathrm{ej}}=1\times10^{-2}\,M_{\odot}$. Since the ejected matter is
moving outward with an average velocity of
$\langle\beta\rangle:=\langle\,v\,\rangle/c\sim0.2-0.3$~\citep{Foucart2015,Lehner2016,Radice2016,Radice2018a,
Sekiguchi2016,Bovard2017},
and this is considerably slower than the relativistic outflow produced by
the explosion. Initially, the matter in/outside the torus is set to have
an azimuthal/zero velocity only.

\begin{figure*}
  \begin{center}
    \includegraphics[width=0.98\columnwidth]{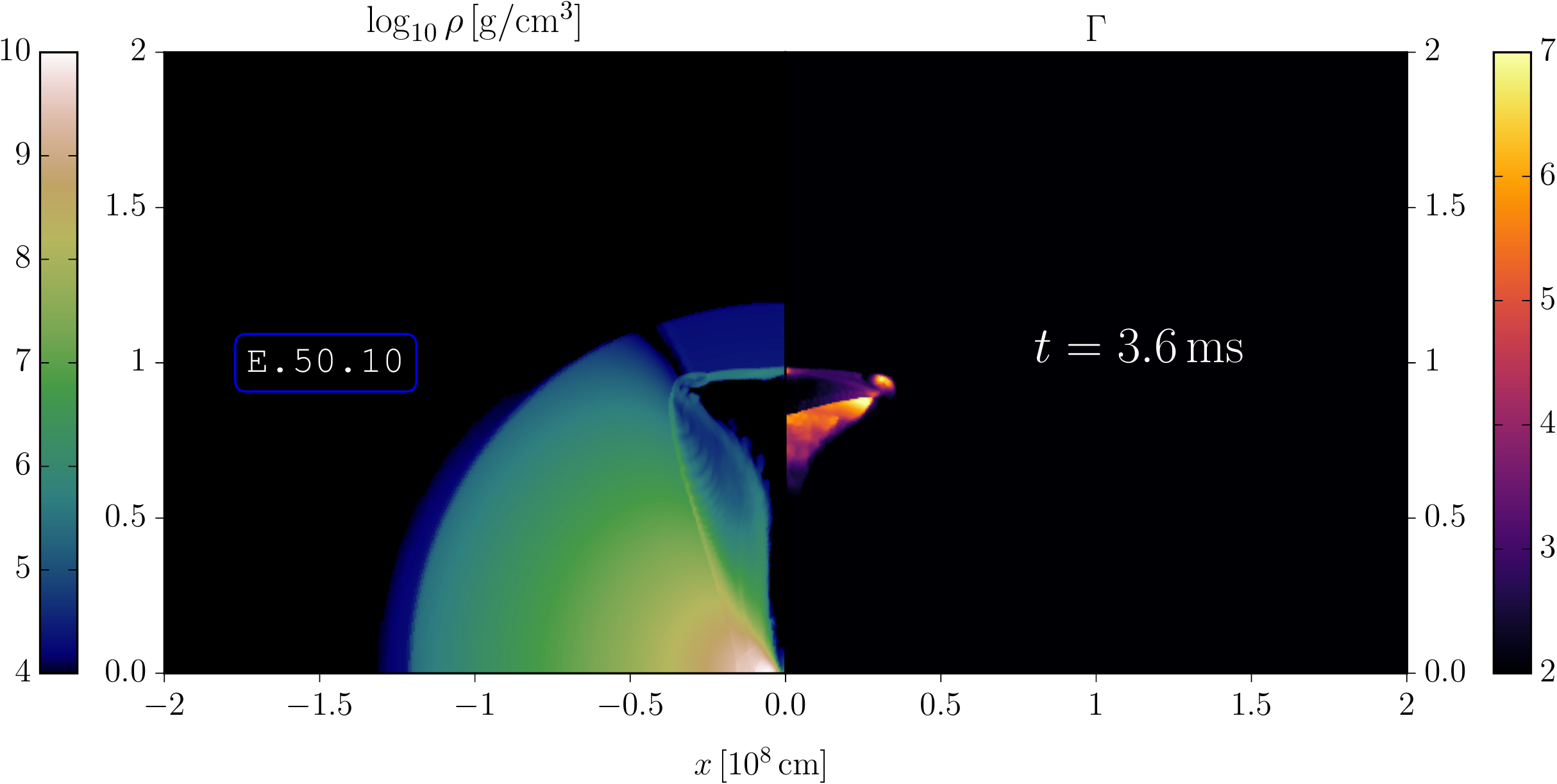}
    \includegraphics[width=0.98\columnwidth]{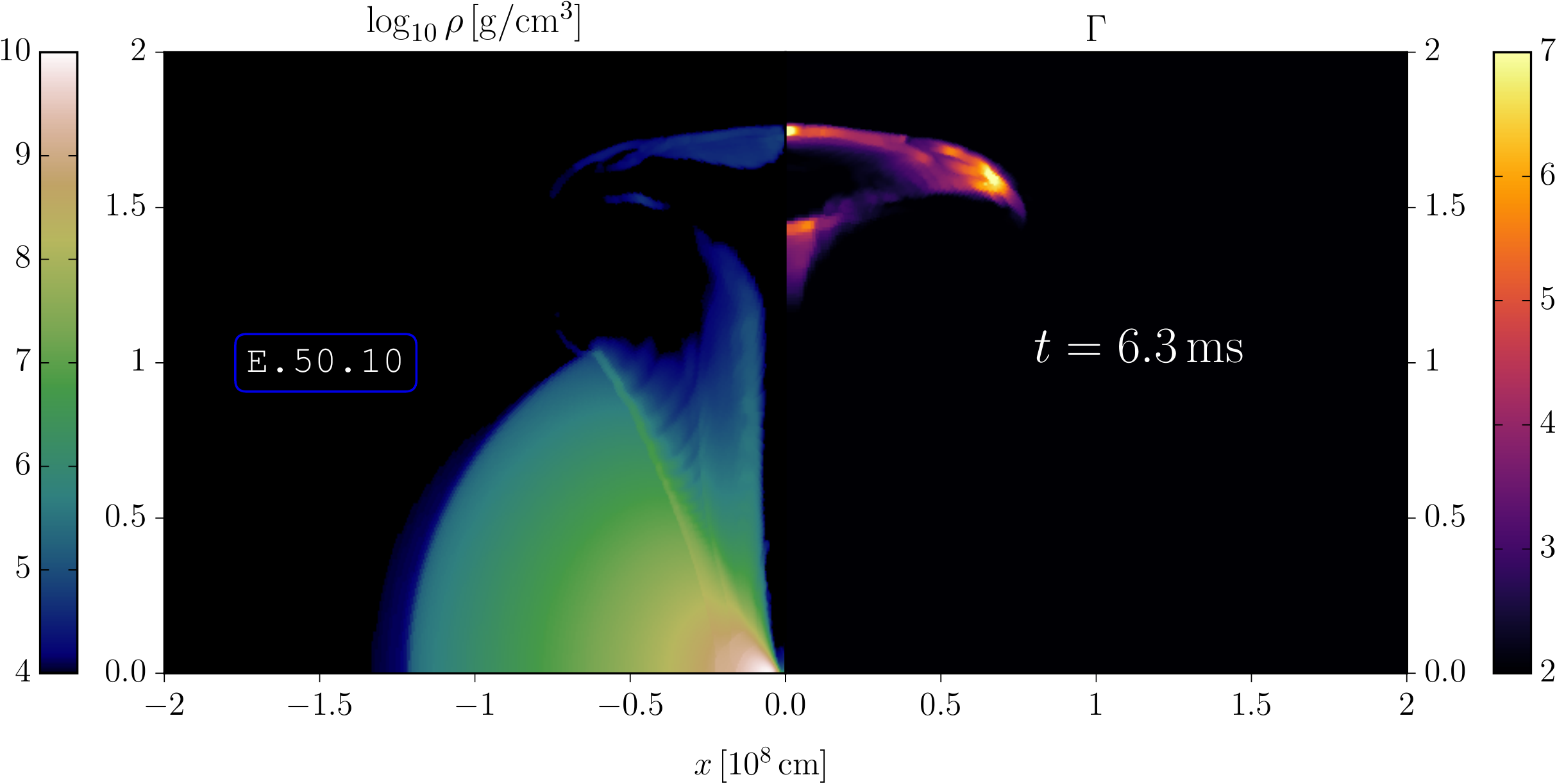}
    \includegraphics[width=0.98\columnwidth]{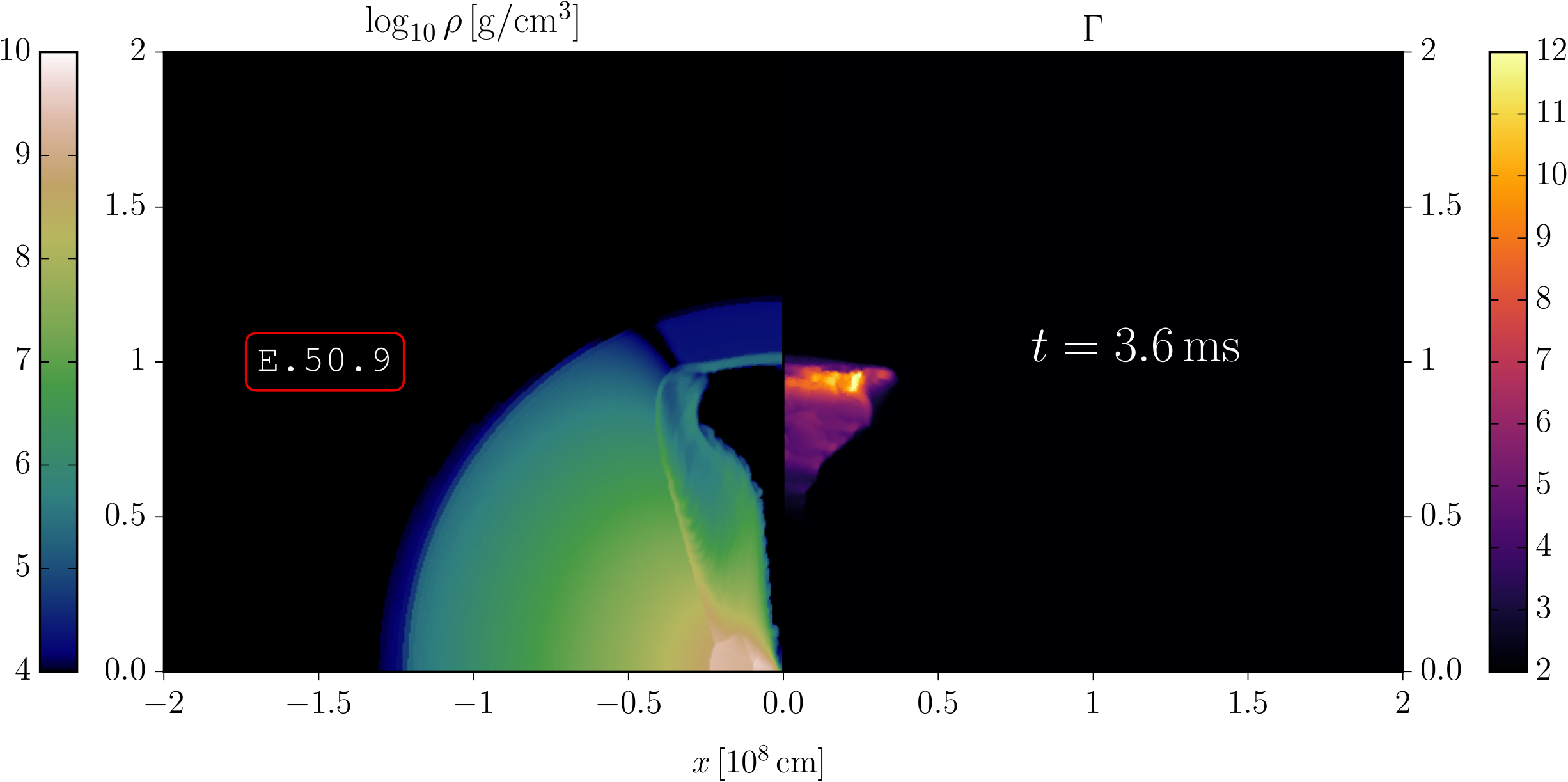}
    \includegraphics[width=0.98\columnwidth]{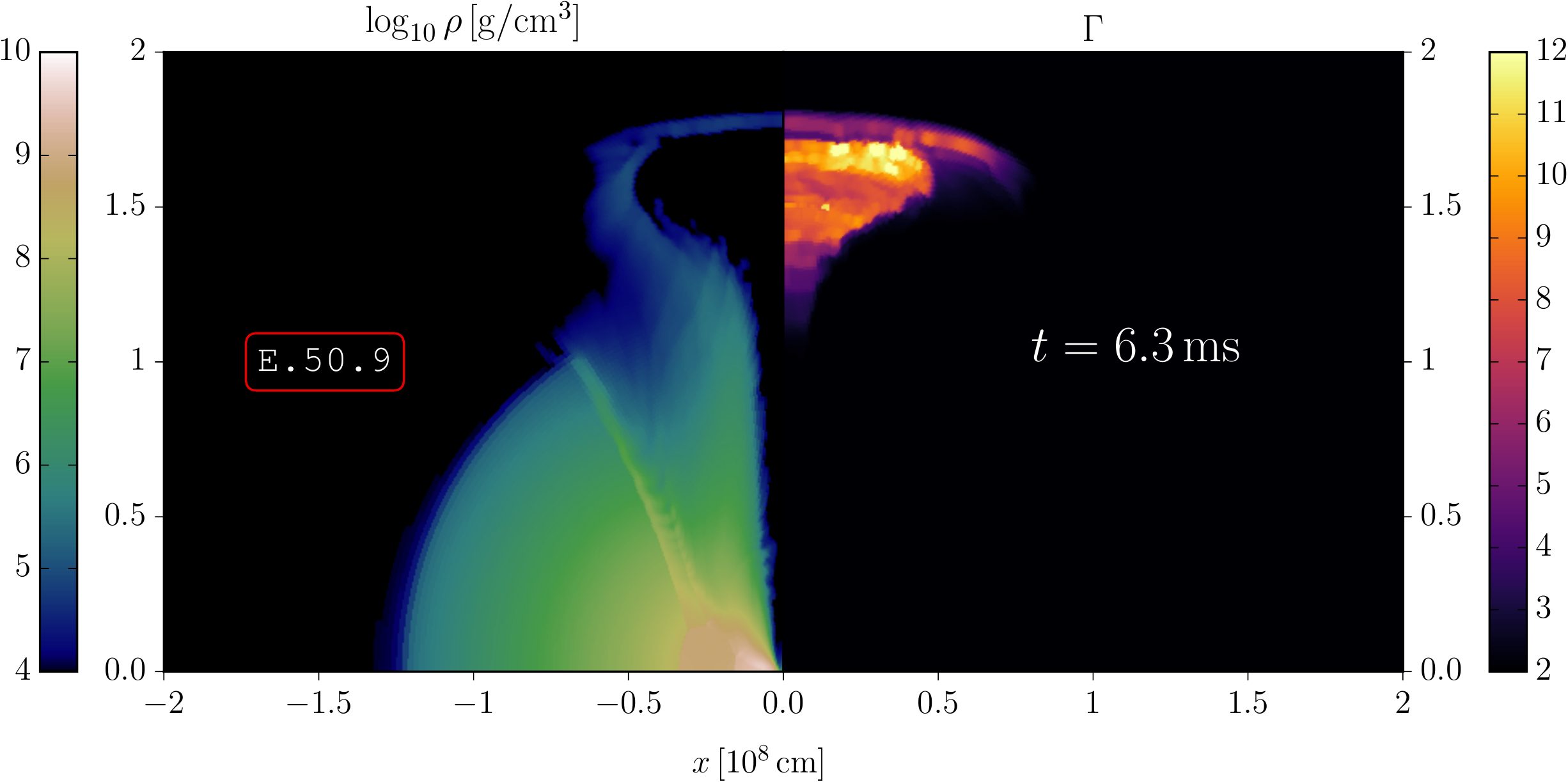}
    \includegraphics[width=0.98\columnwidth]{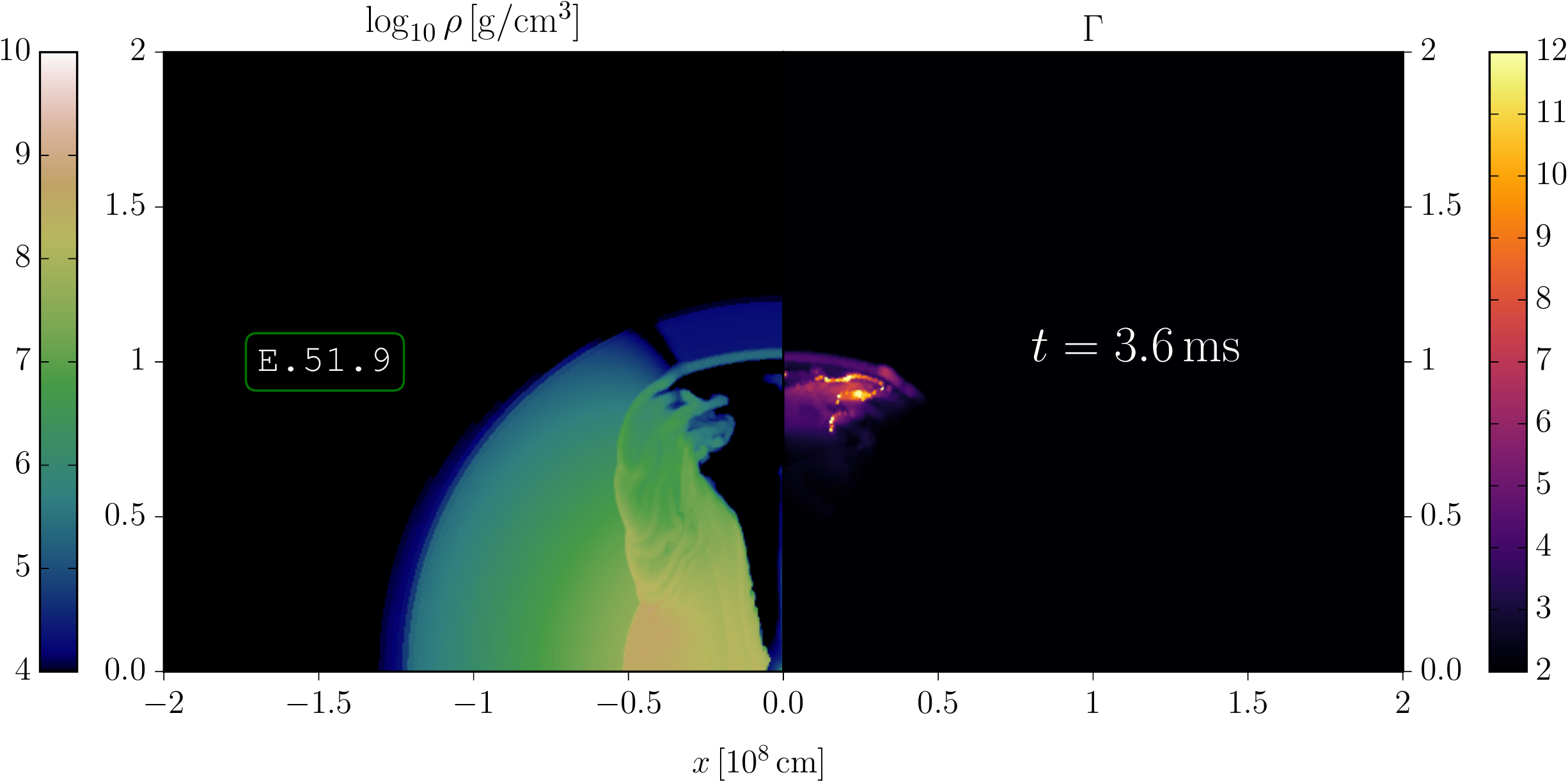}
    \includegraphics[width=0.98\columnwidth]{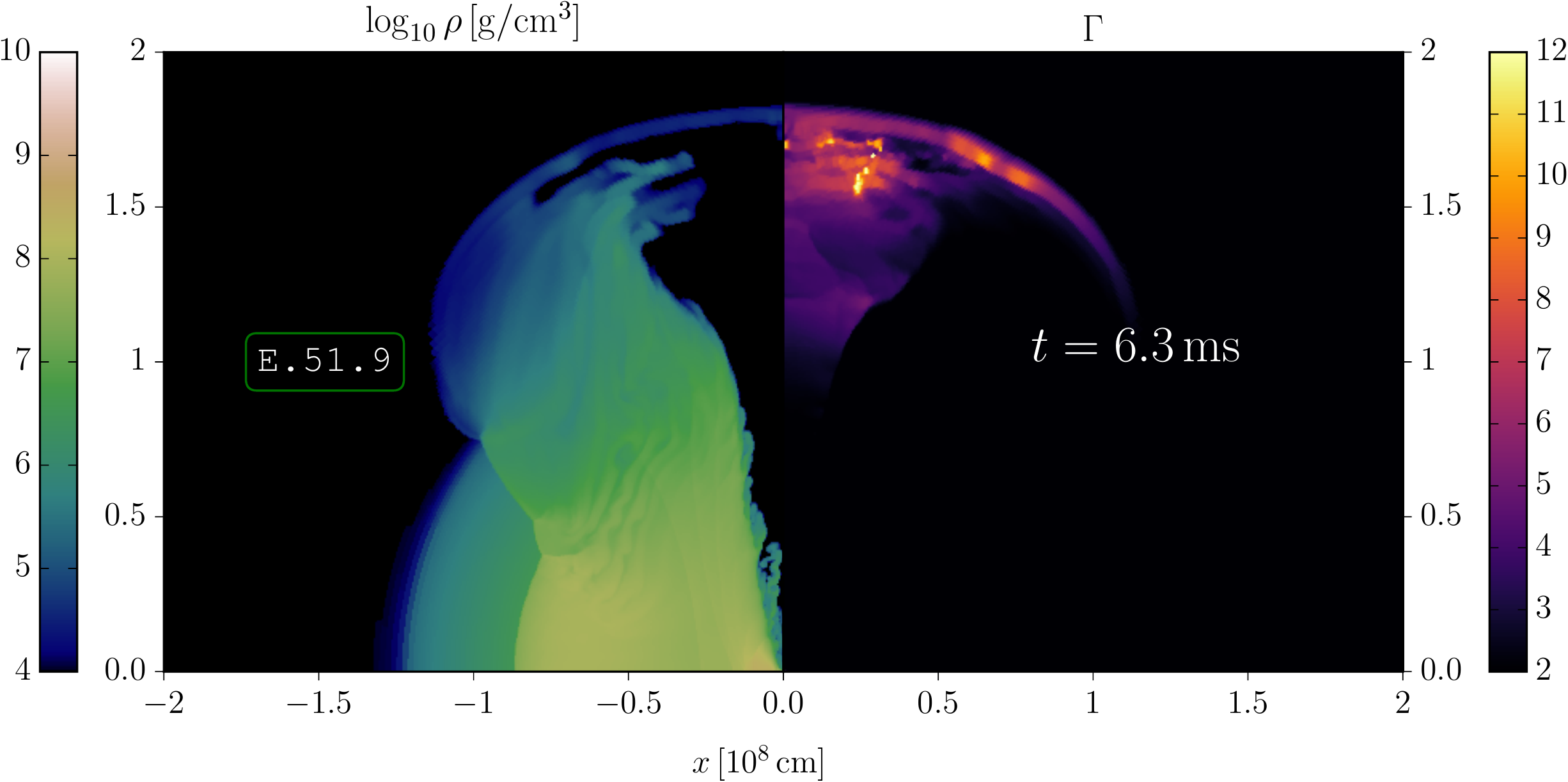}
  \end{center}
  \caption{\textit{Top}: density (left panels) and Lorentz factor (right
    panels) at two different times for an explosion of
    $10^{50}\,\mathrm{erg}$ on a torus of maximum density
    $10^{10}\,\mathrm{g/cm^3}$, \texttt{\small{E.50.10}}. \textit{Middle
      and bottom}: the same, but for \texttt{\small{E.50.9}} and
    \texttt{\small{E.51.9}}, respectively.}
  \label{fig:exp-1}
\end{figure*}

Finally, the amount of energy released needs to be specified. Although
studies of the energetics of collapsing isolated neutron
stars exist ~\citep{Most2017}, the amount of energy available is still
unclear. Numerous numerical simulations have shown that magnetic energy
can be amplified after the merger due to instabilities either at the
shear layer between the two neutron-stars or in the bulk of the
remnant. All in all, the magnetic energy can reach values as high as
$10^{51}\,\mathrm{erg}$~\citep{Kiuchi2017} and even higher values are
thought possible. At the time of collapse of the remnant, most of this
magnetic energy is released almost isotropically and generates a strong
shock with the external matter. In our simulations, which are purely
hydrodynamical, we assume that all of this energy is released in the form
of internal (thermal) energy that we inject at time $t=t_{\mathrm{coll}}$
by initializing an excess internal energy in a spherical shell between
$r_{\mathrm{in}}\simeq20\,\mathrm{km}$ and
$r_{\mathrm{out}}\simeq23.8\,\mathrm{km}$, which is located between black
hole and the torus inner edge. More precisely, if $E_{\mathrm{exp}}$ is
the amount of energy released in a spherical shell of radii
$r_{\mathrm{in}}$ and $r_{\mathrm{out}}$, we inject in the corresponding
cells a pressure
$p_{\mathrm{sh}}\propto9\pi\,E_{\mathrm{exp}}/[\rho_{\mathrm{tor}}(r_{\mathrm{in}}^3-r_{\mathrm{out}}^3)]$.

Although we  neglect magnetic fields here, we also note that the large
majority of the simulations of short-GRB jets discussed recently in the
literature have been performed mostly in the hydrodynamic
limit~\citep{Murguia-Berthier2016,Lazzati2017a,Murguia-Berthier2017,Gottlieb2018,Xie2018},
although some MHD investigations have also been
performed ~\citep{Bromberg2018,Kathirgamaraju2018}. Because of the scale
freedom in the test-fluid approximation, the ratio of the shell energy
density to torus density is the only degree of freedom. Hence,
increasing/decreasing the injected energy together with the density of
the torus leaves the whole setup unchanged. This has the important
advantage that only five simulations need to be performed to cover the
12 representative cases reported in Table~\ref{tab:initial}.

\begin{figure*}
  \begin{center}
    \includegraphics[width=0.4\textwidth]{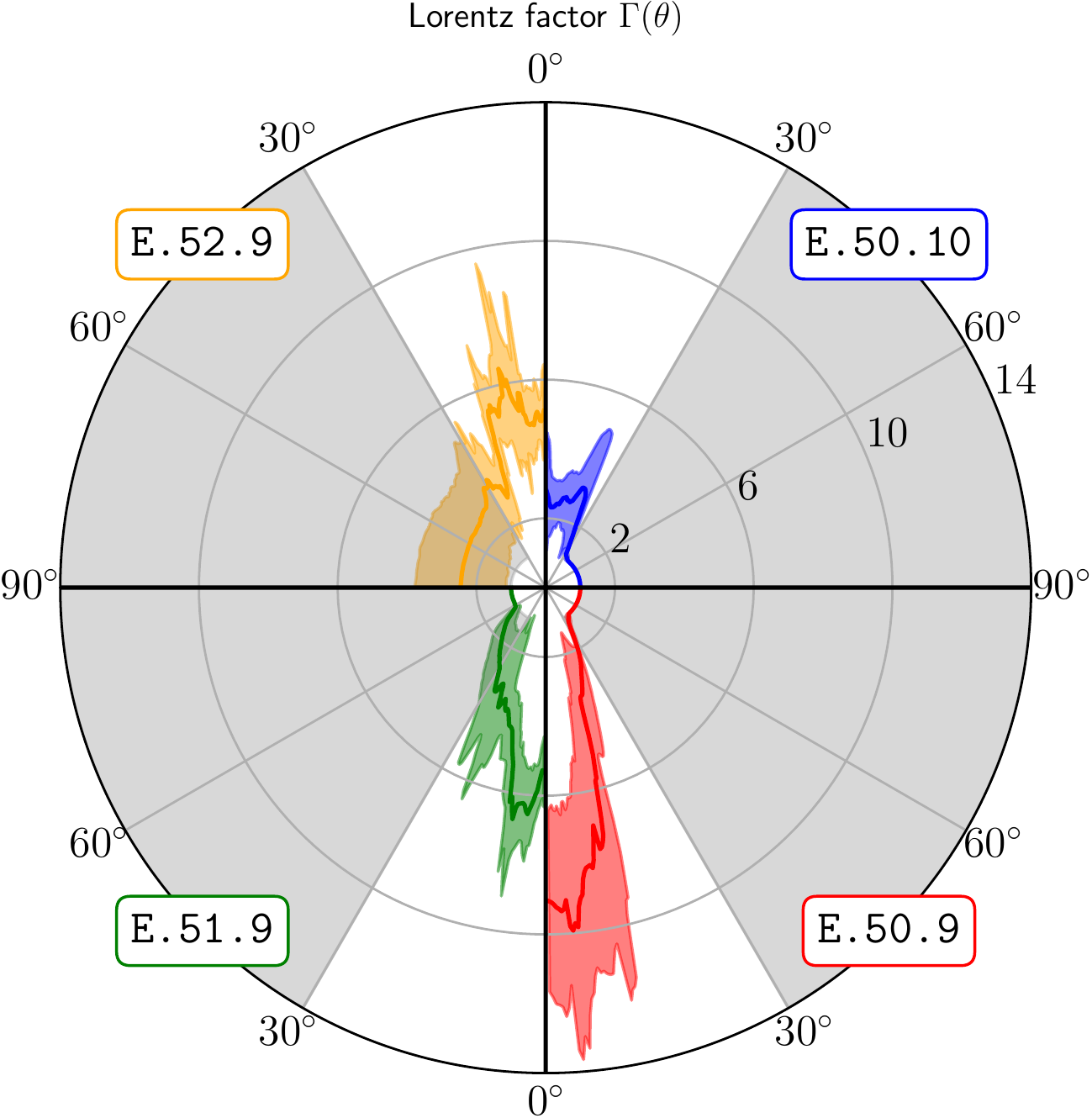}
    \hskip 1.0cm
    \includegraphics[width=0.4\textwidth]{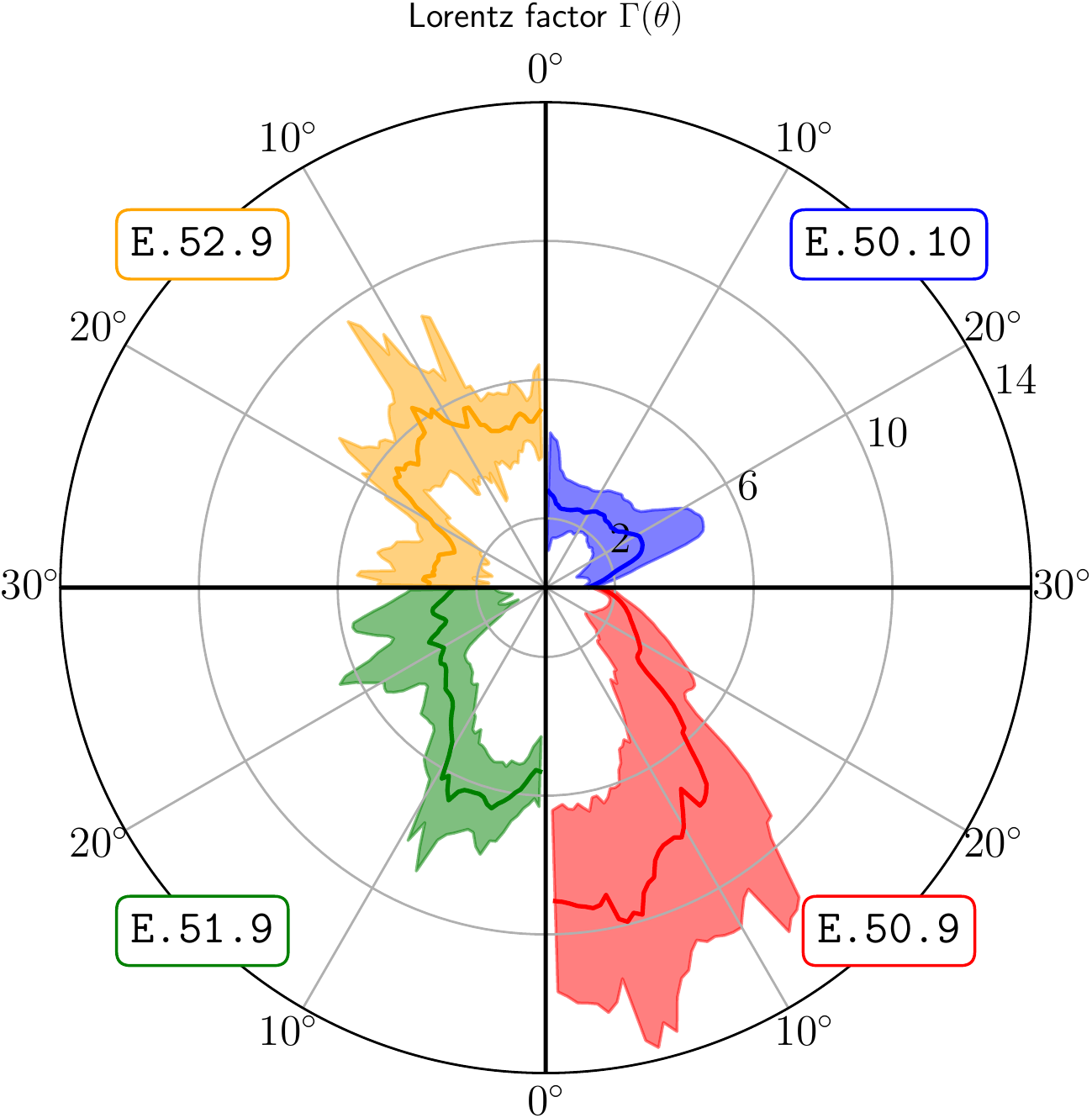}
  \end{center}
  \caption{Polar plots of the Lorentz factor for four representative
    outflows over a quadrant (left panel), or within a cone of
    $30^{\circ}$ (right panel); the thick lines show the average values,
    while the shaded region shows the $1$-$\sigma$ variance.}
  \label{fig:ang-sl}
\end{figure*}

\section{Results} 
\label{sec:model}

As the energy is injected at the center of the system, it produces a
strong shock on the equatorial plane, which however cannot break out
because of the high-density material that it finds on its way. On the
other hand, the less dense regions near the pole allow for matter to
expand rapidly and then break out in a low-density region.

\begin{figure}
  \begin{center}
    \includegraphics[width=0.4\textwidth]{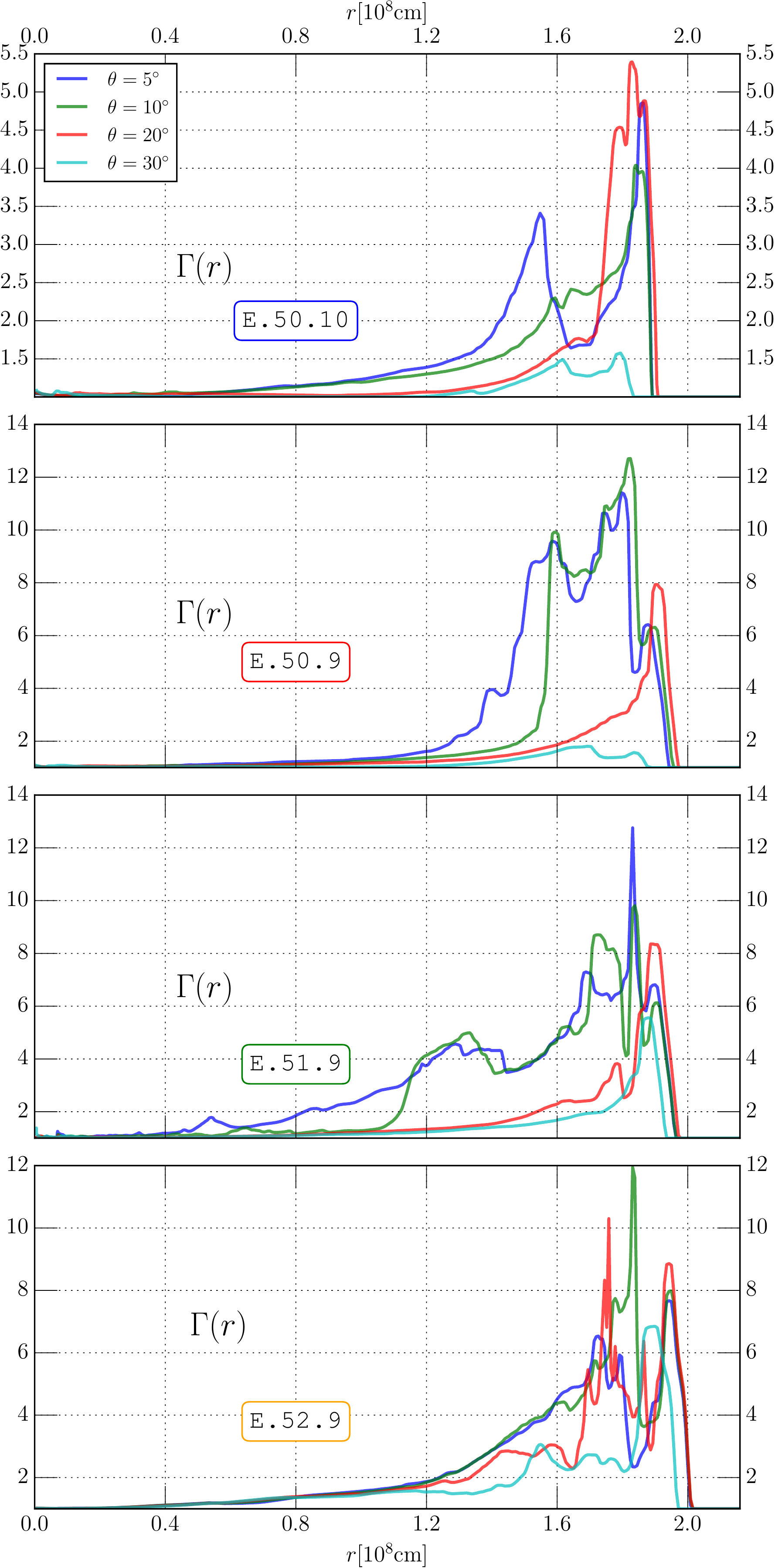}
  \end{center}
  \caption{Radial dependence of the Lorentz factors; each plot shows
    $\Gamma(r)$ along polar slices at
    $\theta=5^{\circ},\,10^{\circ},\,20^{\circ},\,$ and $\,30^{\circ}$.}
  \label{fig:radial}
\end{figure}
\begin{figure*}
  \begin{center}
    \includegraphics[width=1.0\textwidth]{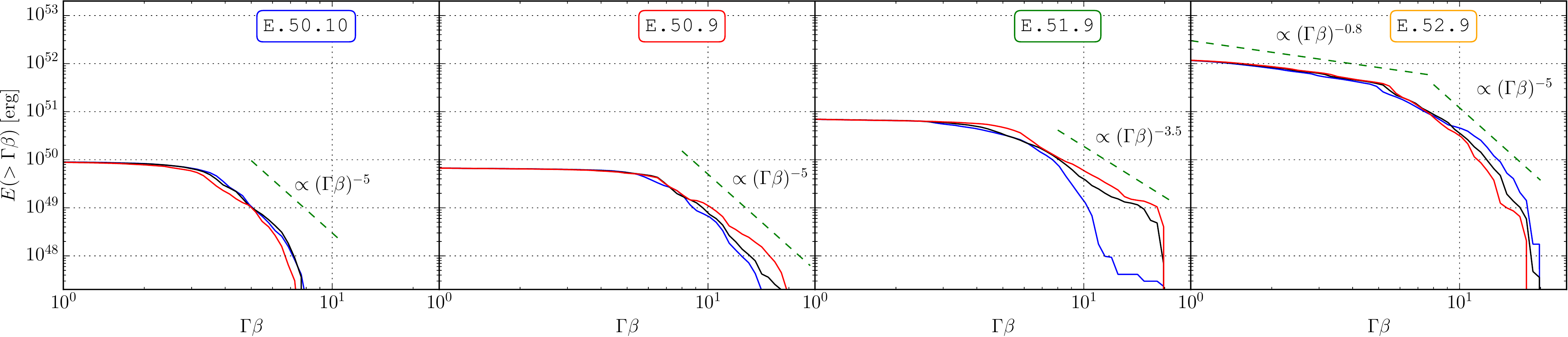}
  \end{center}
  \caption{Energy structure for four representative outflows. The
    blue, black, and red solid lines represent the distribution at
    different times: $t=6.6,\,8.0,\,$ and $9.3\,\mathrm{ms}\,$
    respectively.}
  \label{fig:energy}
\end{figure*}

Rather quickly (see top left panel of Fig.~\ref{fig:exp-1}), the shocked
material in the polar region reaches Lorentz factors higher than five and
acceleration is still ongoing. As time progresses, the shock piles up
matter as it passes through the funnel leaving behind a very low-density
region. In addition, the shock begins a sideways expansion as the outer
parts of the ejected-matter distribution (\ie the torus), decrease in
density (see top right panel of Fig.~\ref{fig:exp-1}). At a radius of
$\sim\,1200\,\mathrm{km}$, the density has already fallen by almost six
orders of magnitude, and can be taken as the break out radius of the
outflow into the low-density region. As the energy released is increased
(middle and bottom panels of Fig.~\ref{fig:exp-1}), it either is converted
into kinetic energy of the outflow, which then acquires larger
Lorentz factors (\eg \texttt{\small{E.50.9}}), or  disrupts the torus
without a significant acceleration of the outflow (\eg
\texttt{\small{E.51.9}}).

Besides what is shown in Fig.~\ref{fig:exp-1}, the angular structure of the
outflow can be best quantified through the polar plots in
Fig.~\ref{fig:ang-sl}, which report the Lorentz factors achieved as
measured in slices of constant radius, \ie $r\sim2000\,\mathrm{km}$, and
integrated over a time interval of
$\tau_{\mathrm{avg}}\sim1.8\,\mathrm{ms}$. More specifically, each panel
refers in its four quadrants to four representative models, \ie
\texttt{\small{E.50.10}}, \texttt{\small{E.50.9}},
\texttt{\small{E.51.9}}, and \texttt{\small{E.52.09}}, each indicated
with a thick line, while the shaded areas show the $1\sigma$ variance
over $\tau_{\mathrm{avg}}$, \ie the $68\%$ variation of the Lorentz
factor at each angle. Furthermore, while the left panel in
Fig.~\ref{fig:ang-sl} shows the global angular structure, \ie across all
latitudes, the right panel zooms in on the polar region, \ie with
$0\leq\theta\leq30^{\circ}$, so as to offer a more detailed
representation of the structure of the outflows.

Starting from the upper-right quadrant of Fig.~\ref{fig:ang-sl}, which
still refers to \texttt{\small{E.50.10}}, it is easy to see that the
Lorentz factor reaches peak values of $\Gamma\simeq2$ at an angle of
$15^{\circ}$-$20^{\circ}$, which is significantly larger than the average in a cone
$0\lesssim\theta\lesssim30^{\circ}$. 
Increasing the energy by one order
of magnitude, as in \texttt{\small{E.50.9}}, (lower-right quadrant), the
average Lorentz factor increases significantly, reaching values
$\Gamma\simeq10$ near the polar axis, while being overall confined within
an angle of $30^{\circ}$. As the energy is further increased by one order
of magnitude, as in \texttt{\small{E.51.9}} (lower-left quadrant), there
is enough energy released to accelerate large amounts of ejecta. As a
result, the outflow has comparable bulk Lorentz factors, but is now
propagating in a much wider funnel, which extends up to
$\theta\lesssim60^{\circ}$. Finally, in the most extreme case of
\texttt{\small{E.52.9}}, matter at essentially all latitudes, hence also
the dense torus on the equatorial plane, is pushed away, acquiring a
Lorentz factors $\Gamma\sim2-6$ at angles as large as
$\theta\sim90^{\circ}$.

In addition to a complex angular structure, the outflow also develops a
nontrivial radial stratification, reported in Fig.~\ref{fig:radial},
where we show $\Gamma(r$) at $t\sim1.8\,\mathrm{ms}$ in four different
directions, \ie $\theta=5^{\circ},\,10^{\circ},\,20^{\circ},$ and
$30^{\circ}$. In all cases, and as expected in a spherical blast wave,
the fast moving parts of the outflow are those in the leading edge of the
wave~\citep{Blandford-McKee1976}, which reach Lorentz factors of the
order $\Gamma\sim5-10$. At the same time, and in contrast to a standard
blast wave, the velocity in the tail of the wave does not decrease
monotonically, but shows secondary peaks that have comparable Lorentz
factors (indeed even larger in the case of \texttt{\small{E.50.10}}), in
particular within a cone of $\sim10^{\circ}$ from the polar axis. This
complex stratification is due to multiple shock reflections in the
low-density funnel and to a shear flow as the wave interacts with the denser
torus material. Interestingly, the conditions reproduced here could lead
to a tangential-velocity booster~\citep{Aloy:2006rd}, which develops when
the dynamics of a relativistic jet can be assimilated to the motion of
two fluids, the inner one being much hotter moving with a large
tangential velocity with respect to the cold, slowly moving outer
fluid. These conditions are often met in numerical simulations of short
GRBs.

With the exception of the top one, all panels in Fig.~\ref{fig:radial}
have the same vertical scale and very comparable values of the Lorentz
factor at least within an angle of $\theta\lesssim20^{\circ}$. Since the
panels report injected energies differing by one order of magnitude (or
more), this similarity indicates that as more energy is released, this is
converted to accelerate larger portions of the ejecta rather than to
further accelerate the material near the polar axis. This is an important
finding pointing out that \emph{wide outflows require energy releases in
  excess of $10^{52}\,\mathrm{erg}$}; of course, such a large amount of
injected magnetic energy is rather unrealistic.

It is also interesting to note the similarities of the models described
here with the properties of the outflows that can fit the late afterglow
of GW170817. The latter can be explained in terms of a wide-angle outflow
that is re-energized by slower parts of the flow following the first and
fast-moving leading part. As the fast part of the flow starts moving into
the interstellar medium (ISM) and decelerates, the slower parts catch up
causing an increase in the afterglow flux~\citep{Troja2018}.

\cite{Mooley2018} have shown that the early radio-afterglow of GW170817
can be explained with a cumulative energy profile of
$E(>\Gamma\beta)\propto(\Gamma\beta)^{-\alpha}$, where $E$ is the sum of
the internal and kinetic energy and $E(>\Gamma\beta)$ is the energy of
the outflow in excess of $\Gamma\beta$; the observations suggest that
$\alpha \simeq 5$. In Fig.~\ref{fig:energy} we show the structure of the
energy as a function of $\Gamma\beta$ for the four
representative models. The energy is measured after the front part of the
outflow has reached the low-density region and has been reported at three
different times with a separation of $1.4\,\mathrm{ms}$ in time. As can
be seen from Fig.~\ref{fig:energy}, the high-velocity tails of all flows
indeed show a steep profile that scales nearly like
$\propto(\Gamma\beta)^{-5}$; furthermore, the knee between the
constant-energy distribution and the power law moves to larger values of
$\Gamma\beta$ as the energy released is increased. Note also that in the
exceptionally energetic model \texttt{\small{E.52.9}}, the slow part of
the flow exhibits another power law with slope $-0.8$.

The shock produced from the interaction between the expanding outflow and
the surrounding ISM emits multi-wavelength synchrotron radiation and is a
helpful tool to distinguish between models~\citep{Nakar2011}. The
afterglow of GW170817 indicated that there was an energy increase at the
shock with the ISM either by a radial stratification of the flow (which
is naturally developed in our models) or by the widening of the beaming
cone of a relativistic jet, which gradually comes into our line of
sight. Clearly, an important development of the results presented here
will be represented by radiative-transfer calculations of the radiation
produced as this outflow impacts on the ISM and leaves a distinctive
imprint, possibly in the degree of polarization, which could break the
degeneracy among different possible models~\citep{Gill2018}.

\section{Conclusions} 
\label{sec:conc}
We have performed a series of general-relativistic hydrodynamical
simulations modeling a possible scenario accompanying the collapse of
the remnant produced by a BNS merger. In particular, we have shown that
if a spherical ``explosion,'' namely, an isotropic and sudden release of
magnetic energy is triggered around one second after the merger, a narrow
outflow with a relativistic core or a wide-angle, mildly relativistic
outflow is generated. The outflow propagates mostly through the polar
regions, where the density is much smaller. Although only mildly
relativistic, the outflow quickly catches-up with the far slower merger
ejecta, breaking-out in the low-density region where it acquires a wide
angular structure. Since our initial energy injection is perfectly
isotropic, the final angular distribution of the outflow and its radial
stratification are the consequence of the propagation through the highly
anisotropic distribution of the ejecta. Interestingly, the outflow
develops an energy dependence close to $\propto\Gamma\beta^{-5}$, in an
encouraging agreement with the recent radiative models of the early
afterglow of~\citet{Mooley2018}.

While this work is meant as the exploration of a plausible scenario for
GRB 170817A, several future improvements can be made. First, while
axisymmetry is probably a good approximation, it is important to validate
these results with three-dimensional simulations. Second, magnetic fields
will be present inside and outside of the merger remnant and with a complex
topology ~\citep{Siegel2014}. The interaction of the ejecta with these
magnetic fields needs to be taken into account with MHD
simulations. Thirdly, when the black hole is formed, it will ring down,
producing pulses of electromagnetic radiation~\citep{Most2017}, possibly
impacting the stratification of the outflow. Finally, we note that our
model should not be seen as being in contrast with the standard jet-formation
scenario in short GRBs. Rather, it argues that not all BNS mergers can
produce a jet and shows that even without the production of a collimated
jet, a mildly relativistic outflow can be expected if sufficient magnetic
energy is released when the remnant collapses.

\medskip
It is a pleasure to thank D. Giannios and R. Gill for useful
discussions. A.N. is supported by a von-Humboldt Fellowship.  Support also
comes from the ERC synergy grant ``BlackHoleCam: Imaging the Event
Horizon of Black Holes'' (grant No. 610058), ``PHAROS,'' COST Action
CA16214; the LOEWE-Program in HIC for FAIR; the European Union's Horizon
2020 Research and Innovation Programme (Grant 671698) (call
FETHPC-1-2014, project ExaHyPE).  The simulations were performed on the
clusters SuperMUC (LRZ, Garching), LOEWE (CSC, Frankfurt), and HazelHen
(HLRS,Stuttgart).

\bibliographystyle{yahapj}

\end{document}